\documentclass[aps,pre,preprint,nofootinbib,eqsecnum,tightenlines,
showpacs,amsmath,titlepage]{revtex4}
\usepackage{graphicx}
\begin{document}

\title{On the Temperature Dependence of the Casimir Effect}

\author{I. Brevik}
\email{iver.h.brevik@ntnu.no} 
\author{J. B. Aarseth}
\email{jan.b.aarseth@ntnu.no}
\affiliation{Department of Energy and Process Engineering, Norwegian University
of Science and Technology, N-7491 Trondheim, Norway}

\author{J. S. H{\o}ye}
\email{johan.hoye@phys.ntnu.no}
\affiliation{Department of Physics, Norwegian University of Science and
Technology, N-7491 Trondheim, Norway}

\author{K. A. Milton}
\email{milton@nhn.ou.edu}
\affiliation{Oklahoma Center for High Energy Physics and
Department of Physics and Astronomy, The University of Oklahoma,
Norman, Oklahoma 73019, USA}
\date{\today}

\begin{abstract}
The temperature dependence of the Casimir 
force between a real metallic plate and a
metallic sphere is analyzed on the basis of optical data concerning the
dispersion relation of metals such as gold and copper.  
Realistic permittivities
imply, together with basic thermodynamic considerations, that the
transverse electric zero mode does not contribute.  This results in observable
differences with the conventional prediction, which does not take this
physical requirement into account.  The results are shown to be consistent
with the third law of thermodynamics, as well as 
being not inconsistent with current experiments.
However, the predicted temperature dependence should be detectable in
future experiments.  The inadequacies of approaches based on {\it ad hoc\/}
 assumptions,
such as the plasma dispersion relation and the use of surface impedance without
transverse momentum dependence, are discussed.
\end{abstract}

\maketitle

\section{Introduction}
\label{intro}
There are many corrections that one in principle has to
take into account when calculating the Casimir force between two
bodies; the corrections may come from finite temperatures, finite
extensions of the plates used in the experiments, corrugations on
the plates, etc. (Recent reviews of the Casimir effect can be
found in Refs.~\cite{milton04,milton01,bordag01}.) The 
correction that we will be concerned with in the present paper is
the one coming from finite temperatures. For the most part, we
will consider the temperature dependent Casimir force between a
compact sphere of radius $R$  and a plane substrate. The sphere is
situated at a fixed distance $a$  from the plane ($a$ denotes the
minimum distance between the surfaces). The sphere and the
substrate are assumed nonmagnetic, but we consider the  case where
they may be made from different materials. We will moreover assume
the proximity force theorem \cite{blocki77} to hold; this means
that $a$ must be much less than $R$.
(For corrections to this see Refs.~\cite{jaffe04,Gies,Emig}.)
 On experimental grounds it is
evidently desirable to calculate the Casimir forces in a realistic
way. We will here take advantage of the excellent numerical
dispersive data to which we have access for the materials gold and
copper (and also aluminum) (courtesy of Astrid Lambrecht and Serge
Reynaud).  We know how the permittivity
$\varepsilon(i\zeta)$ varies with imaginary frequency $\zeta$ over
seven decades, $\zeta \in [10^{11}, 10^{18}]$ rad/s. We use these data
to calculate the forces at two different temperatures, namely at
room temperature, $T=300$ K, and at $T=1$ K. The latter
temperature is conveniently attainable numerically, and it can for all 
practical purposes be identified with zero temperature. (The lowest 
temperature that we actually tested was $T=0.2$ K. If T becomes lower, 
we leave  the   frequency domain for our numerical dispersion data. 
It turns out that there are  very small deviations between calculated 
values of the force for $T=1$ K and for $T=0.2$ K.)  
 We obtain in this way a realistic picture of the
finite temperature correction for these materials.

It ought to be emphasized that we are not adopting the so-called
modified ideal metal (MIM) model, which assumes unit reflection coefficients
for all but the transverse electric
(TE) zero mode [see Eq.~(\ref{10}) below]; 
rather, we are using real data together
with the assertion (based on thermodynamical and electrodynamical
arguments) that the TE zero mode is absent, 
as that is an isolated point
which cannot be extracted from data alone.  Our approach is that which we have
followed in other recent
papers \cite{milton04,hoye03,brevik04}. The absence of a TE zero mode
contribution to the Casimir effect for a real metal was
discussed in detail in Ref.~\cite{hoye03}, and also in
Ref.~\cite{sernelius04}. The result of this assumption, for instance, 
in the MIM model is the
presence of a linear temperature term in the expression for the
Casimir force between two planes, in the limit where $aT \ll 1$.
The calculation for a real metal yields a linear temperature
correction for low, but not too low temperatures, so that for very low
temperatures the force and the free energy have zero slope.
By contrast, in the conventional (old) model for an ideal metal
(IM) the TE zero mode is included, and it implies that this
linear temperature term is omitted. We ought to stress
here that at $T=0$ the mentioned difference between a MIM and an
IM model goes away, as the contributions from the zero frequency
TE mode as well as from the zero frequency TM mode become buried
in an integral over imaginary frequencies from zero to infinity.

The experiments of immediate interest for the present
theory are the atomic force microscopy  (AFM) tests, performed in
particular by Mohideen {\it et al.}~\cite{mohideen98}. A point that we
ought to emphasize  here is that  previous analyses have most
likely overestimated the accuracy of the AFM experiments. Thus the
recent paper of Chen {\it et al.}~\cite{chen04}, which is based
upon a reanalysis of the experiment of Harris {\it et al.}~(listed
in \cite{mohideen98}), claims an over-all experimental precision
to be  at the 1 percent level. In that apparatus a gold-coated
polystyrene sphere mounted on a cantilever of an AFM was brought
close to a metallic surface and the deflection was measured.
However, as discussed in Refs.~\cite{iannuzzi04,milton04}, at the
very short distance of 62 nm (the minimum distance) the force at
$a=62 \,\mbox{nm}+\delta$ differs from the force at $a=62$ nm by more than
3.5 pN (the experimental uncertainty claimed by the authors) when
$\delta$ is larger than a few angstroms. This means that $a$
should have been measured with atomic precision in order to
correspond to the accuracy claimed. As for temperature
corrections, these were found in \cite{chen04} to be negligible,
this being related to their acceptance of the plasma dispersion
relation for the material.

As the absence of the zero frequency TE mode has been controversial,
we give in Sec.~\ref{sec4}  a discussion of this in view of 
 recent work by Bezerra {\it et al.}~\cite{bezerra04}
They argue that this mode should be present. In Sec.~\ref{sec5} 
we give additional
support to our arguments by showing that for a pair of anisotropic polarizable 
particles the Casimir force can vanish in certain directions as
the temperature increases towards $\infty$, and although there are regions
of negative entropy connected with the Casimir effect, 
there is no indication that 
thermodynamics is violated. Violation of thermodynamics is used as an 
argument by Bezerra {\it et al.}~\cite{bezerra04} to require the presence of the 
TE zero mode.  The inadequacy of using a surface impedance approach without
including transverse momentum dependence is briefly reviewed in Sec.~\ref{si}.

 In this paper we put $\hbar =c=k_B=1$.

\section{General Formalism and Dispersive Properties}
\label{sec2}

Let the sphere of radius $R$ be nonmagnetic, and have a
permittivity $\varepsilon_1$. As mentioned, the sphere is situated
a distance $a$ above a plane substrate; we let the nonmagnetic
substrate have permittivity $\varepsilon_2$. According to the
proximity force theorem \cite{blocki77} the attractive force
$\cal{F}$ between sphere and plane at temperature $T$ can in the
limit $a/R \ll 1$ be given approximatively as the circumference of
the sphere times the surface free energy density $F$ in the
parallel-plate configuration: ${\cal{F}}=2\pi R F(a)$. Following
essentially the notation of Ref.~\cite{hoye03} we can then write
the force as
\begin{equation}
{\cal{F}}=\frac{R}{\beta a^2}{\sum_{m=0}^\infty}^\prime\int_{m
\gamma}^\infty y\,dy \left[ \ln
(1-\Delta_1^{\rm TM}\Delta_2^{\rm TM}e^{-2y})+\ln(1-\Delta_1^{\rm TE}\Delta_2^{\rm TE}e^{-2y})\right].
\label{1}
\end{equation}
Here $y$ is the dimensionless quantity $y=qa$, and
\begin{equation}
q=\sqrt{k_\perp^2+\zeta_m^2},\quad\zeta_m=2\pi m/\beta,\quad
\gamma=2\pi
 a/\beta , \label{2}
\end{equation}
${\bf k_\perp}$ being the component of $\bf k$ parallel to the
plates in the parallel-plate configuration. Further, $\zeta_m$
with $\beta=1/T$ are the Matsubara frequencies, and $\gamma$ is
the dimensionless temperature. Superscripts TM and TE in
Eq.~(\ref{1}) refer to the transverse magnetic and electric modes.
The prime on the sum means that the zero mode has to be counted
with half weight.  With the conventional Lifshitz variables
defined as
\begin{equation}
s=\sqrt{\varepsilon -1+p^2},\quad p=q/\zeta_m, \label{3}
\end{equation}
we define the two kinds of $\Delta$'s, the reflection coefficients for
a single interface, as
\begin{equation}
\Delta^{\rm TM}=\frac{\varepsilon p-s}{\varepsilon p+s},\quad
\Delta^{\rm TE}=\frac{s-p}{s+p} \label{4}
\end{equation}
for each medium 1 and 2, respectively. If the two media are equal,
$\Delta_1=\Delta_2$ for each kind of mode, then
\begin{equation}
(\Delta^{\rm TM})^2 \equiv A_m,\quad (\Delta^{\rm TE})^2 \equiv B_m,
\label{5}
\end{equation}
where $A_m,\,B_m$ are the TM, TE coefficients defined in
Ref.~\cite{hoye03}. Note that
$s_1=\sqrt{\varepsilon_1-1+p^2},\,s_2=\sqrt{\varepsilon_2-1+p^2}$,
with $p=q/\zeta_m$ being the same quantity in the two cases. The
permittivities $\varepsilon(i\zeta_m)$ are functions of the
imaginary Matsubara frequencies $\zeta_m$. In the general case
where the media are dispersive, $\Delta^{\rm TM}$ and $\Delta^{\rm TE}$
depend both on $p$ and on the Matsubara integer $m$. If the media are
nondispersive, $\Delta^{\rm TM}$ and $\Delta^{\rm TE}$ are functions of
$p$ only, independent of $m$.

 As a general warning, we mention that the proximity theorem assumed here
 requires $a/R$ to be very small.  Thus, within the framework of
 the optical path method recently considered by Jaffe and
 Scardicchio \cite{jaffe04}, believed to be more robust than the
 proximity approximation, disagreement with the latter
 approximation was found already when $a/R$ became larger than a
 few percent, whereas the method they propose agrees accurately
 with the recent exact numerical result of Gies {\it et al.}~\cite{Gies}.

\subsection{Dispersive properties}
\label{sec2.1}
As mentioned in the Introduction, we will use accurate numerical
data for the variation of $\varepsilon$ with frequency for two
different substances: gold and copper. These data
refer to room temperature measurements. For gold, the data are
shown graphically in Refs.~\cite{lambrecht00,hoye03}. For
frequencies up to about $1.5\times 10^{15}$ rad/s (note that 1
eV=$1.519\times 10^{15}$ rad/s), the data are nicely reproduced by
the Drude dispersion relation
\begin{equation}
\varepsilon(i\zeta)=1+\frac{\omega_p^2}{\zeta(\zeta+\nu)},
\label{6}
\end{equation}
where for gold the plasma frequency is $\omega_p=9.0$ eV and the
relaxation frequency $\nu=$ 35 meV. For $\zeta > 2\times 10^{15}$
rad/s the Drude curve however lies below the experimental curve.

All the dispersive data, of which we are aware, refer to room
temperature. Now, as we will be interested in the Casimir force
also at low temperatures, we are faced with the problem of how to
estimate the permittivity $\varepsilon(i\zeta,T)$ under such
circumstances. Numerical trials indicate rather generally that the
Casimir force is very robust under  variations in the input
values for the permittivity, but at least this issue is a matter
of principle. One possible way to proceed is to write the
permittivity as
\begin{equation}
\varepsilon(i\zeta, T)=1+\frac{\omega_p^2}{\zeta[\zeta+\nu(T)]},
\label{7}
\end{equation}
and make use of the Bloch-Gr\"{u}neisen formula for the
temperature dependence of the electrical resistivity $\rho$. This
problem was discussed in Appendix D of Ref.~\cite{hoye03}. One may
in this way estimate the temperature relaxation frequency to be, in eV,
\begin{equation}
\nu(T)=0.0847
\left(\frac{T}{\Theta}\right)^5\int_0^{\Theta/T}\frac{x^5e^xdx}{(e^x-1)^2},
\label{8}
\end{equation}
where $\Theta=175$ K for gold. 
This formula implies that $\varepsilon(i\zeta, T)$ is somewhat
higher for low $T$ than for $T=300$ K (cf. Fig.~1 in
\cite{hoye03}), when the frequencies are less than about $10^{14}$
rad/s. It is instructive to compare with the parallel-plate
configuration, where it is known that the most important
frequencies for the Casimir force are lying in the region $\zeta a
\sim 1$. For $a=1\,\mu$m it corresponds to $\zeta \approx
2.5\times 10^{14}$ rad/s. For smaller gap widths---where the
metallic properties of the medium fade away and its plasma
properties become more dominant---it follows that the most
important frequencies become higher. Taking  all things together,
we expect that the influence from the temperature variation in
$\nu(T)$ is rather small. Some numerical trials that we have done
support this expectation.

Another important point to be mentioned here is that the
Bloch-Gr\"{u}neisen argument sketched above neglects the effect
from impurities. These give rise to a nonzero resistivity at zero
temperature \cite{khoshenevisan79}. This fact strengthens our
argument for setting  the contribution from the
Casimir force from the  TE zero mode for a metal equal to zero. The issue has
been considered in detail also by Sernelius and Bostr{\"o}m
\cite{sernelius04,bostrom00}. As one can see from Fig.~2 in
\cite{sernelius04}, there exists a temperature dependent
contribution to the Casimir force from the TE modes. This
temperature dependence occurs for low frequencies at low
temperature and extends to higher frequencies with increasing
temperature. However, the temperature influence fades away before
the the first nonzero Matsubara frequency is reached. It is therefore
permissible to neglect the temperature dependence in $\nu(T)$.
What remains important in $\nu$ is the constant term
$\nu(T=0)\ne0$ that is
due to elastic scattering.  The consequence is that 
\begin{equation}
\zeta^2[\varepsilon(i\zeta)-1]\to 0,\quad\mbox{as}\quad \zeta\to0,
\end{equation}
so from (\ref{4}) $\Delta^{\rm TE}$ vanishes at $m=0$.
 If one neglects this crucial constant $\nu(T=0)$,
one can end up with a violation of the Nernst heat theorem
\cite{bezerra02}.  See also Bostr\"om and Sernelius \cite{bostrom04}
for related discussion of these points.

Therefore, in the following we will restrict ourselves to using the
room-temperature values for $\nu$ throughout, even when
calculating ${\cal{F}}(T)$ at different temperatures. Even with
this simplification, we point out  that the temperature dependence
in the Casimir force turns up in a rather complex way. Namely, the
temperature occurs at three different places:

(i)  in the prefactor in Eq.~(\ref{1});

(ii) in the lower limit of the integral;

(iii) in the dependence of $\Delta^{\rm TM,TE}_1$ and
$\Delta_2^{\rm TM,TE}$ on $T$ via the Matsubara frequencies in the
permittivity: $\varepsilon = \varepsilon(i2\pi mT)$.

\section{Numerical Results}
\label{sec3}
\begin{figure}
\centering
\includegraphics[height=10cm]{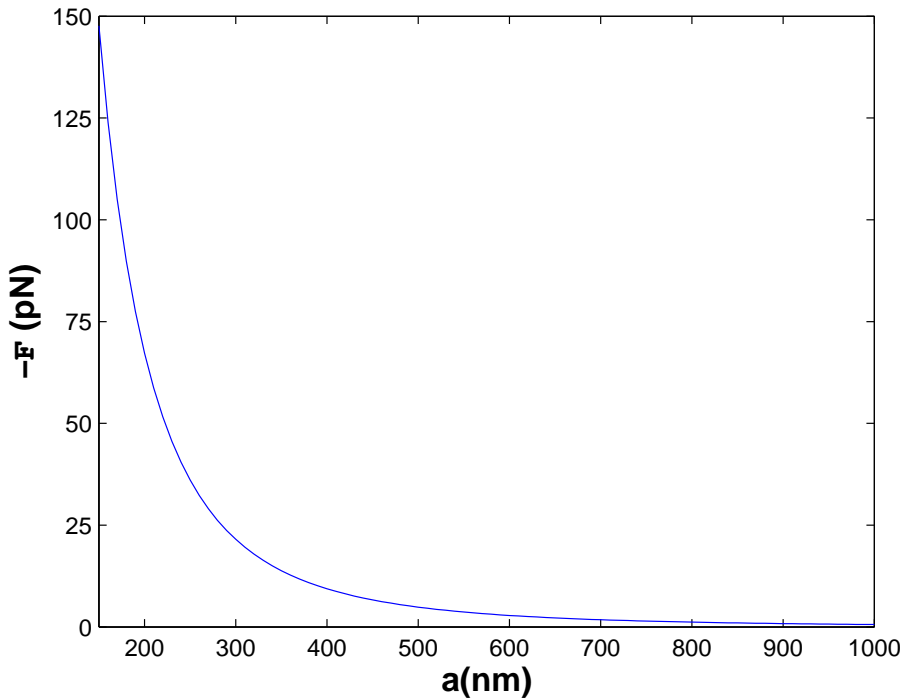}
\caption{Force ${\cal F}$ between a gold sphere and a gold
plate versus gap $a$, when $T=300$ K and $R=296$ $\mu$m.}
\label{fig1}
\end{figure}

Figure \ref{fig1} shows, for a gold sphere and a gold plate, how the
attractive force ${\cal F}(a)$ varies with $a$ in the interval
from about 150 nm to 1 $\mu$m, when $R=296$ $\mu$m and $T=300$ K. As mentioned, the
empirical data for $\varepsilon(i\zeta)$ are directly usable as
input in Eq.~(\ref{1}). When $a=200$ nm, the force is calculated
to be 67.22 pN.

A similar calculation can be made for a very low temperature, in
order to show the magnitude of the temperature influence in the
force. We assume throughout, as mentioned earlier, that $\nu= 35$
meV.  Using MATLAB we found  $T=1$ K  to be a numerically stable
and reasonable lower limit. This temperature is moreover low
enough to be identifiable with $T=0$ for all practical purposes.

We ought to stress that we choose to perform the $T \approx 0$
calculation {\it numerically}, inserting realistic data for
$\varepsilon(i\zeta)$. This is in principle different from the
conventional $T=0$ calculation for an idealized metal, where one
simply puts $\varepsilon =\infty$ for all frequencies. (Recall
that at $T=0$ the difference between a MIM and an IM model goes
away, because of the very close spacing between the Matsubara
frequencies.)

\begin{figure}
\centering
\includegraphics[height=10cm]{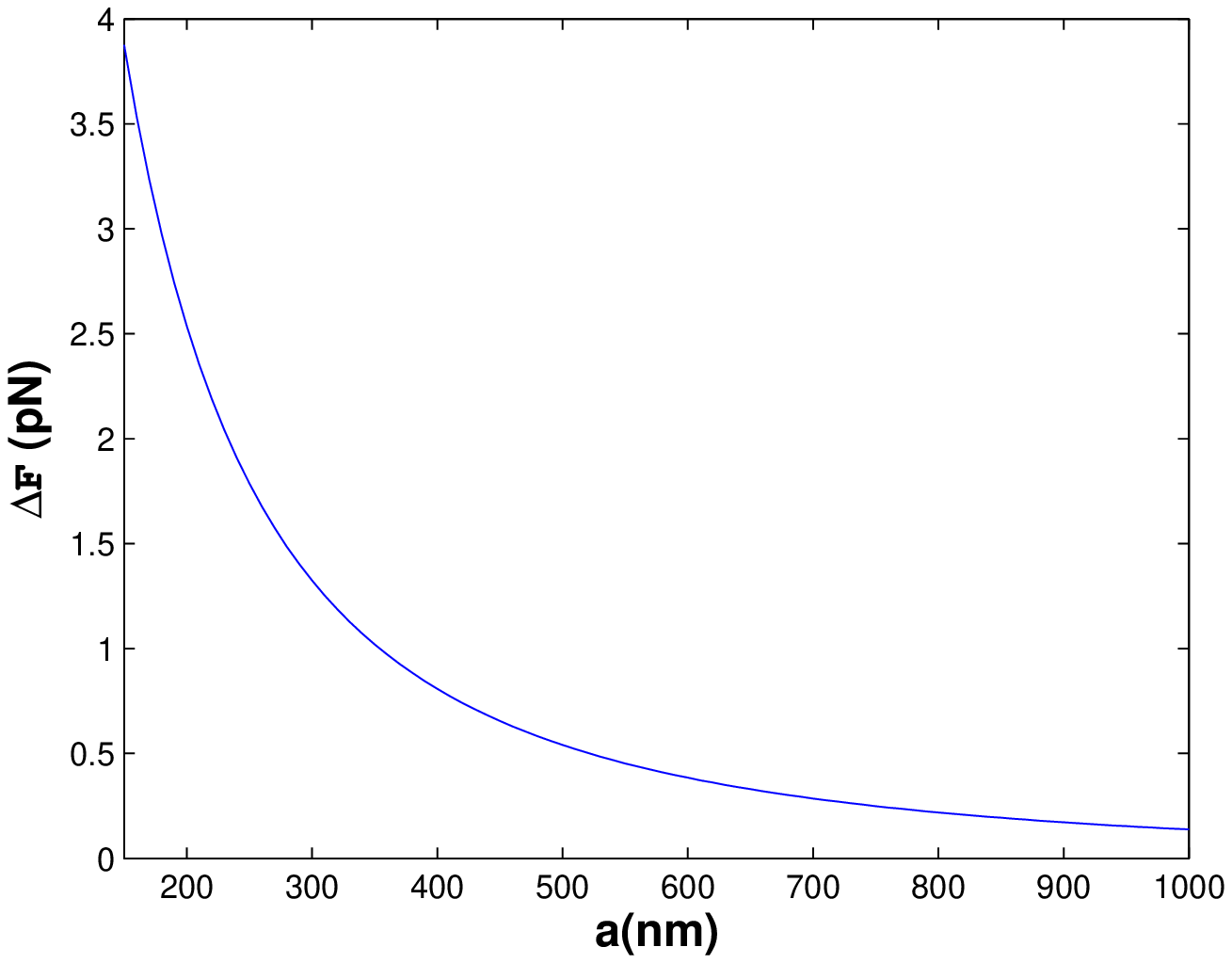}
\caption{Force difference $\Delta {\cal F}=|{\cal
F}\,(1\,\rm{K})|-|{\cal F}\,(300 \,{\rm K})|$ between a gold sphere
and a gold plate, versus gap $a$ for $R=296$ $\mu$m.}
\label{fig2}
\end{figure}

Rather than showing the calculated result for ${\cal F}(1\,{\rm
K})$ explicitly, we show  in Fig.~\ref{fig2} the difference between the
forces, $\Delta {\cal F}$, defined as
\begin{equation}
\Delta {\cal F}={\cal F}\,(300 \,{\rm K})-{\cal F}\,(1\,\rm{K})
=|{\cal F}\,(1\,\rm{K})|-|{\cal F}\,(300 \,{\rm K})|. \label{9}
\end{equation}
An important property seen from this curve is that $\Delta{\cal F}$ is
positive. The force is thus {\it weaker} at room temperature than
at $T=0$. This is the same effect as was found in Fig.~5 in
\cite{hoye03}. This behavior is thus a consequence of Lifshitz'
formula plus realistic input data for the permittivity; there are
no further assumptions involved. When $a=200$ nm, we find $\Delta
{\cal F}$= 2.54 pN, which means that the force is reduced by 3.6
percent compared to the $T=0$ case. 

For larger distances, $a=400$ nm, the temperature effect becomes
larger. Thus for $a=400$ nm the force is 9.38 pN at $T=300$ K  and
10.19 pN at $T=1$ K, yielding  a 7.9 percent reduction at room
temperature. At $a=1\; \mu$m, the corresponding numbers are 0.59
pN at $T=300$ K and 0.73 pN at $T=1$ K, which means a 19 percent
reduction. This agrees in magnitude well with the temperature
corrections in the case of parallel-plate geometry, as is seen
from Fig.~5 in Ref.~\cite{hoye03}.

Admitting an error of $10^{-8}$
in the $m$ summation in Eq.~(\ref{1}), we found the necessary
number of terms to be in excess of 34000 in the case of the lowest
separation investigated numerically, $a=50$ nm (not shown in the
figure). When $a=200$ nm, about 11000 terms were required. At
larger separations the necessary number of terms became considerably
reduced; thus the case $a=1\,\mu$m corresponded to about 2700
terms.

The calculation of the force between a gold sphere and a copper
plate gave very similar results. Thus for $a=200$ nm the force was
67.19 pN at $T=300$ K and 69.75 pN at $T=1$ K, corresponding to a
reduction of 3.7 percent at room temperature.  At $a=1\, \mu$m,
the forces turned out to be the same (to the accuracy of two
decimals) as in the Au-Au case.

\begin{figure}
\centering
\includegraphics[height=10cm]{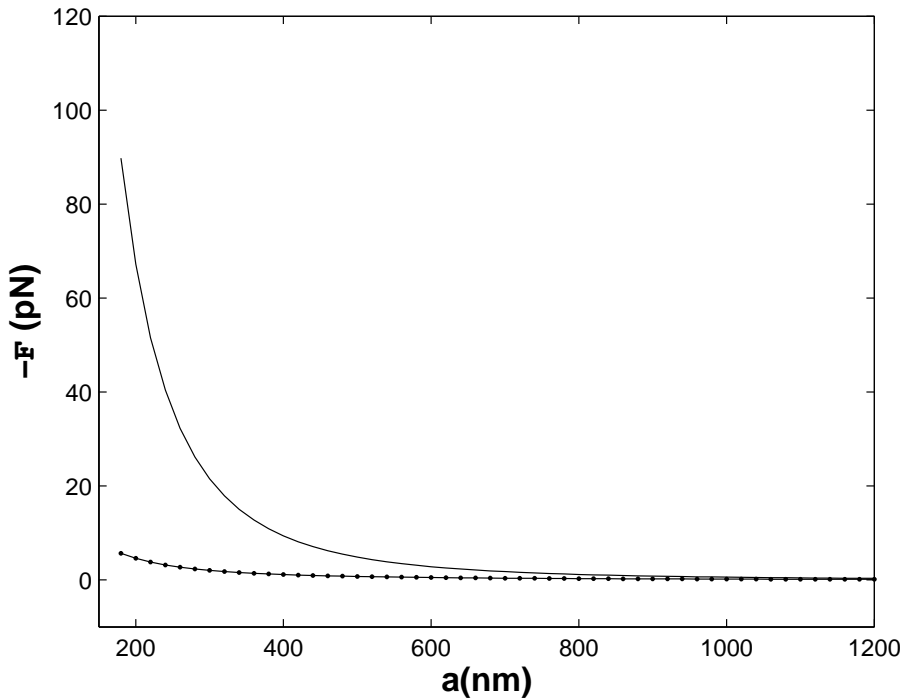}
\caption{Force ${\cal F}$ between a gold sphere and a copper
plate versus $a$, when $T=300$ K and $R=296$ $\mu$m. Bottom curve is the $m=0$
contribution. 
}
\label{fig3}
\end{figure}

In Fig.~\ref{fig3} we show, by the full line, how the force between a gold
sphere and a copper plate varies with $a$, at $T=300$ K. The curve
is calculated from Eq.~(\ref{1}), using the empirical data for
these two materials directly. Our reason for giving this curve
anew, in spite of its identity with the  curve in Fig.~\ref{fig1} for all
practical purposes, is that we have supplied the following
new element, namely
at the bottom of the figure we show explicitly
the contribution from the $m=0$ term to the force. Namely, for the
MIM model in which $A_0=1,\,B_0=0,\,A_m=B_m=1$ for $m \ge 1$
\cite{hoye03}, we have for the free energy
\begin{equation}
\beta F^{\rm MIM}=\frac{1}{4\pi}\int_0^\infty q\,dq
\ln(1-e^{-2qa})+\frac{1}{\pi}\sum_{m=1}^\infty
\int_{\zeta_m}^\infty q\,dq\ln (1-e^{-2qa}), \label{10}
\end{equation}
showing the $m=0$ contribution in the first term. Thus
\begin{equation}
\beta F^{\rm MIM}(m=0)=-\frac{\zeta(3)}{16\pi a^2}, \label{11}
\end{equation}
meaning that the corresponding force contribution $2\pi R
F^{\rm MIM}(m=0)$ becomes
\begin{equation}
{\cal F}^{\rm MIM}(m=0)=-\frac{\zeta(3)}{8}\frac{R}{\beta a^2}.
\label{12}
\end{equation}
From the figure this term contributes an increasing part of the total 
force for increasing  $a$, and gives the full contribution in the classical 
limit $a \rightarrow \infty$.

We would like to make a direct comparison of our prediction
 with the recent experiment of Decca
{\it et al.}~\cite{decca03,decca03a}. This experiment measured the
Casimir force between a gold sphere and a copper plate by means of
a microelectromechanical torsional oscillator, for separations in
the range 0.2--2 $\mu$m. The radius of the sphere was 
$296\pm2$ $\mu$m, thus the same radius as we have assumed above. Their
measured values are shown, for instance, in Fig.~3 in
Ref.~\cite{decca03}. 
However, the scale of their figure for the total force prevents comparison
with our numerical results with sufficient accuracy to draw firm conclusions
about the magnitude of the temperature dependence.

But Fig.~\ref{fig3} in Ref.~\cite{decca03} also gives a comparison with theoretical
values by plotting the difference. These theoretical values are evaluated at 
$T=0$, but as we do, they also use a Drude model to obtain the dielectric 
constant in the limit of low frequencies. Then they find a small temperature 
correction with the same sign as we have theoretically predicted.

From Fig.~\ref{fig3} in Ref.~\cite{decca03} we estimate $\Delta \cal F$ to be 
around 1 pN for $a=200$ nm, and to be  too small to be discernible for large 
gaps, $a \ge 600$ nm (the scatter of the experimental points is considerable). 
This can be compared with our Fig.~\ref{fig2}, where we calculated 
$\Delta \cal F$ to be 2.56 pN for $a=200$ nm and 0.14 pN for $a=1\,\mu$m. 
This reasonably good agreement between experimental and theoretical results is 
encouraging, and it indicates that our finite temperature calculations are on 
the right track.

We should also mention that Ref.~\cite{decca03a} refers to dynamical 
measurements 
that are claimed to rule out our results.  However, we believe that the
systematic theoretical and experimental uncertainties in this measurement
are larger than estimated by those authors.  In particular,
 we re-emphasize the uncertainty of measuring the Casimir 
force due to uncertainties in estimating a systematic shift of position as 
earlier discussed by Iannuzzi {\it et al.}~\cite{iannuzzi04} and 
Milton \cite{milton04}, as mentioned in the Introduction.

\section{Comment upon the Transverse Electric Zero Mode}
\label{sec4} 
Bezerra {\it et al.}~\cite{bezerra04}
 claim that the Drude dielectric function for metals cannot be used in the 
 theory for the Casimir force.
Their opinion is that it violates the Nernst theorem in thermodynamics and 
is furthermore ruled out by a recent experiment. Instead the
plasma relation should be used for the dielectric function. The latter
implies the presence of a transverse electric mode at zero frequency 
besides the static dipole-dipole interaction; the latter being 
the zero-frequency limit of the transverse magnetic mode. Including 
such an transverse electric mode adds a term linear in 
temperature to the Casimir 
force by which it is increased by a factor of two in the high 
temperature limit.

These authors point to the standard theory,
sketched above in Sec.~\ref{sec2.1}
 where the relaxation parameter $\nu$ varies with temperature $T$ and 
vanishes at $T=0$. This is then a situation where in principle a 
 transverse electric zero mode might be present in the Casimir force
although it should not be present according to Maxwell's equations of 
electromagnetism. However, its presence for $\nu=0$, but not otherwise, 
would make physical phenomena discontinuous as then $\nu=0$ will give 
something different from taking the limit $\nu\rightarrow 0$. Also we 
question whether a possible vanishing of the relaxation parameter 
at $T=0$ can dictate the behavior for
$T>0$ where $\nu>0$. (As a remark we here note that strictly speaking the 
statistical mechanical derivation is exact only for dielectric functions 
independent of temperature, i.e., when induced dipole moments are harmonic 
oscillators. A $T$-dependence reflects anharmonicity. But we do not expect 
that this is of crucial significance here.)

As a result, Bezerra {\it et al.}~\cite{bezerra04}
 conclude that the Drude dielectric function 
is thermodynamically inconsistent and cannot be used to calculate the thermal 
Casimir force for metals. However, we on the contrary have shown that it is 
thermodynamically consistent leading to zero entropy at $T=0$ in accordance 
with the Nernst theorem \cite{hoye03}. Thus, for  $T>0$ a negative entropy 
contribution due to electromagnetic interaction between media is allowed as 
long as the total entropy is positive. We have earlier studied a 
simplified model in this regard \cite{hoye03}.  The model consists of three 
harmonic oscillators. Two of them are the analogues of the two media that 
interact via the electromagnetic field represented by the third oscillator. 

In the beginning of their Sec.~III, Bezerra {\it et al.}~\cite{bezerra04}
state that our derivations were 
based upon a constant relaxation parameter $\nu$. However, we did not 
require such a limitation, only that $\nu(0)$ is finite. And as the authors 
further note, the value of $\nu$ is commonly very small, but nevertheless 
finite, at $T=0$ due to impurities, 
and as a result the entropy becomes zero at 
$T=0$ as required. Then they write that the negative entropy we found at 
$T=5\times 10^{-4}$\,K is a violation of the Nernst theorem. 
But such a negative 
perturbing entropy is not a violation of the Nernst theorem since $T$ is 
finite as already discussed above. The further discussion about relaxation 
time, impurities, and surface impedance does not invalidate the use of a 
finite $\nu(0)$.

These authors further remark that we consider nonzero wave vectors 
$\bf k_{\bot}$ for which the reflection coefficient $r^2_{\bot} (0,k_{\bot})
=0$. But this does not mean that reflection properties are different for the 
fluctuation field compared to real photons as we deal with only one such 
quantity; and as discussed above, thermodynamics is not violated.
(See Sec.~\ref{si} below for further discussion of $\mathbf{k_\perp}$ 
dependence.)

They further note that a material with dielectric constant 
$\varepsilon=100$, like a metal, gives a Casimir force that decreases with 
temperature in some interval and thus implies a negative entropy contribution.
 They conclude that such a material cannot exist as it would violate 
thermodynamics. Then they claim that real media with such large $\varepsilon$ 
are commonly polar for which $\varepsilon$ rapidly reduces to its optical 
value connected to its electronic polarizability. 
So with plate (or plate-sphere) 
separation of $1\,\mu{\rm m}$ the Casimir force again becomes monotonically 
increasing with $T$. However, this does not preserve 
the monotonic character of the Casimir force in general 
because one can just increase the plate separation. 
Non-monotonic behavior will 
then reappear when this separation exceeds a wavelength corresponding to the 
relaxation frequency (where $\varepsilon$ decreases rapidly).

In Sec.~IV of their paper, they  again conclude that the use of
 the Drude dielectric 
function in the Lifshitz formula violates the third law of 
thermodynamics (the Nernst heat theorem). This conclusion, stated as a 
rigorous proof, is made on the basis that the relaxation parameter will be 
much less than the Matsubara frequencies. But this is not a rigorous proof as 
the relaxation parameter will violate this inequality sufficiently close to 
$T=0$ (assuming $\nu(0)>0$). Furthermore the Drude dielectric function does 
not predict a linear temperature correction  to the Casimir force all the
way down to zero temperature.
As we have shown earlier the Casimir force flattens out and becomes
independent of temperature at $T=0$ in accordance with thermodynamics
\cite{hoye03,brevik04}. 
(But the sharpness of this flattening increases with decreasing  $\nu$.)

We further remark that the use of the Drude dielectric function is consistent
 with the $\varepsilon\rightarrow\infty$ limit (for large separation of 
the plates). Use of the plasma model ($\nu=0$), however, will give a 
discontinuous jump of the force in this limit. Such a jump is not expected 
for a continuous change in a physical parameter. Also for real metals the 
$\varepsilon$ effectively will be finite due to the finite size of the 
plane-sphere configuration. (Here one can note that a medium consisting of 
separate metal spheres of finite size will be like a polarizable medium with 
finite polarizability and dielectric constant.)

Finally the authors in Ref.~\cite{bezerra04} conclude that use of the Drude 
dielectric function (\ref{6}) is in contradiction with experiment. 
However, so far as 
we can see, experiments performed at a single temperature are at present too 
uncertain to draw conclusions about temperature variations. This seems even 
more evident from Ref.~\cite{chen04}.
There detailed analysis of experimental uncertainties are performed and 
various corrections for very short separations are estimated. They find an 
uncertainty of 1.75\,\% at 95\,\% confidence level for 62\,nm separation.
This uncertainty increases to 37.3\,\% for 200\,nm. Earlier it has been 
remarked by others \cite{iannuzzi04} that such experiments are very sensitive 
to accurate determinations of plate separation since the plate-sphere Casimir 
force is proportional to the inverse cube of separation distance. 
However, the authors 
assert that they avoid this problem by making a least squares fit of the 
resulting data by which zero separation is pinpointed with an uncertainty of 
0.15\,nm. But in view of the uncertainties of the experiment this does not 
seem to resolve the disputed temperature dependence as heavy weight is put on 
the shortest separation of 62\,nm where the force is by far the 
largest and changes most rapidly. 
Thus high apparent precision can be obtained for this separation (1.75\,\%) 
while for larger separations the uncertainties are rapidly increasing 
until they
become larger than the magnitude of the disputed thermal effect. 
According to the authors of Ref.~\cite{chen04} 
the thermal effect in dispute is about 1-2\,\% for 62\,nm.

Very recently, a paper has appeared \cite{valeri} giving the microscopic
theory of the Casimir effect.  These authors convincingly demonstrate that
the TE zero mode cannot contribute, although the plasma model gives such a
contribution.
\section{Anisotropic Particles with Negative Casimir Entropy}
\label{sec5}

As mentioned above the appearance of a 
negative Casimir entropy for metals in a region of 
nonzero temperatures has been disputed with the claim that it violates the 
Nernst theorem of thermodynamics \cite{bezerra04}.  However, as we argue this 
negative entropy region does not imply violation of thermodynamics 
since the Casimir 
free energy is a perturbing one and is not the total free energy of two 
interacting systems. Many such examples are known in statistical mechanics,
including that of three interacting oscillators discussed in Ref.~\cite{hoye03}.
To illustrate this point, we here will consider a pair of 
particles with strong anisotropic polarizability and thus polarizable only in 
the $z$-direction, e.g., they may be metal needles. The result for a pair of 
particles with isotropic polarizability is well known and was rederived in a 
novel way by Brevik and H\o ye \cite{brevik88}. 

The latter derivation is easily generalized to anisotropic particles. Using 
the path integral formalism the dipole-dipole interaction of the radiation 
field is given by Eq.~(I5.9). (Here and below the numeral I refers the the 
equations of the reference mentioned.) Equation (I5.9) gives the interaction
energy
\begin{equation}
\beta\Phi(12)=\eta\sum\limits_{n=-\infty}^\infty [\psi_{D\zeta_n}(r)\,
D_{\zeta_n} (12)
+\psi_{\Delta \zeta_n}(r)\,\Delta_{\zeta_n} (12)]a_{1\zeta_n}a_{2\zeta_n}^*
\label{h1}
\end{equation}
with
\begin{equation}
D_{\zeta_n} (12)=3({\bf\hat{r}\cdot\hat{a}}_{1\zeta_n})({\bf\hat{r}\cdot
\hat{a}}_{2\zeta_n}^*)
-{\bf\hat{a}}_{1\zeta_n}\cdot{\bf\hat{a}}_{2\zeta_n}^*,
\end{equation}
\begin{equation}
\Delta_{\zeta_n} (12)={\bf\hat{a}}_{1\zeta_n}\cdot{\bf\hat{a}}_{2\zeta_n}^*
\end{equation}
where the Matsubara frequency $\zeta_n=2\pi n/\beta$. 
The dipole radiating fields $\psi$ are given in (I5.10).
Hats denote unit vectors. The 
${\bf\hat{a}}_{\zeta_n}$ are unit vectors of the Fourier-transformed 
fluctuating 
dipole moments along the ``polymer" path and $\eta\rightarrow 0$ is the 
discretized step length along the ``polymers"
that represent quantized particles in the path integral formalism. 

The Casimir free energy times $-2\beta$ is now obtained from the average of 
expression (\ref{h1}) squared in accordance with (I3.5). For the isotropic 
case one has the average
 $\eta\langle {\bf a}_{\zeta_n}{\bf a}_{\zeta_n}^*\rangle =3\alpha_{\zeta_n}$ 
 where $\alpha_{\zeta_n}$ 
is the polarizability. This follows from Eq.~(I5.3). For the strongly 
anisotropic situation to be considered below one likewise will find
($z$ denotes the $z$-component)
\begin{equation}
\eta\langle a_{z\zeta_n}a_{z\zeta_n}^*\rangle =\alpha_{z\zeta_n} 
\end{equation}
as the only nonzero average as polarizations in $x$- and $y$-directions are 
zero in the present case. 

Furthermore let the positions of the particles relative to each other be such 
that the $z$-component of ${\bf\hat{r}}$ equals $1/\sqrt{3}$. With this the 
$D_{\zeta_n} (12)$ will vanish as polarizations are present only in the $z$-direction.
 (With this relative position the corresponding electric field is transverse 
to the $z$-direction and thus to each of the dipole moments, i.e., there is 
no interaction connected to the $D_{\zeta_n} (12)$-term.) Thus only the 
$\Delta_{\zeta_n} (12)$-term remains. So with polarizations restricted to the 
$z$-direction Eq.~(I5.14) turns into
\begin{equation}
-\beta F=\frac{1}{2}\sum\limits_n\alpha_{z\zeta_n}^2\psi_{\Delta \zeta_n}^2(r)
\end{equation}
or with (I5.10) inserted the free energy is
\begin{equation}
F=-\frac{1}{\beta r^6}\sum\limits_{n=-\infty}^\infty\left(\frac{2}{3}\tau^2
\right)^2 e^{-2\tau}\alpha_{z\zeta_n}^2
\label{h6}
\end{equation}
where 
\begin{equation}
\tau=\frac{2\pi r}{\beta\hbar c}|n|
\end{equation}
as given by (I5.12).

As usual Eq.~(\ref{h6}) gives a negative free energy. However, in the 
classical limit $T\rightarrow\infty$ the Casimir free energy (\ref{h6}), 
and thus the corresponding force, both vanish  since only the 
$n=0$ term will contribute. Furthermore with this free energy the 
contribution to the entropy must be negative (or at least mainly negative) 
as $S=-\partial F/\partial T$, because $F$ must have a generally positive
slope, but $S=0$ at $T=0$ as it should.

\section{Surface Impedance}
\label{si}
Most recently, Mostepanenko {\it et al.}~\cite{geyer03,bezerra04}, 
apparently conceding that their arguments
favoring the plasma model over the Drude model for the dispersion relation
for real metals could not be supported either thermodynamically, 
electrodynamically, or experimentally, have asserted that for real metals
one should use in the reflection coefficients in the Lifshitz formula
not the bulk dielectric permittivity but the surface impedance.
Indeed there is much to be said for using the latter.  However, in may
be shown in general that there is in fact no difference between the
reflection coefficients computed using either description \cite{brevik04}.
There is a one-to-one correspondence between the permittivity 
$\varepsilon$ and the
surface impedance $Z$, which is given by the ratio of the transverse electric
and magnetic fields at the surface.  This correspondence, however, necessitates
in general that both quantities possess dependence on the transverse
momentum $\bf k_\perp$.  As optical data strongly suggest that this
dependence is usually negligible in the permittivity, a strong dependence
on $\bf k_\perp$ is required in $Z$ \cite{brevik04}
\begin{equation}
Z^{\rm TE}(\zeta,k_\perp)=-\frac{\zeta}{\sqrt{\zeta^2\varepsilon(i\zeta)
+k_\perp^2}}.
\end{equation}
The vanishing of $1+Zq/\zeta$ at $\zeta=0$ demonstrates again that the TE zero
mode does not contribute to the Casimir force.  
(Note that this vanishing happens in the Drude,
but not the plasma model.)   In contrast, Ref.~\cite{geyer03,bezerra04} 
completely disregard this transverse momentum dependence and moreover
make an {\it ad hoc\/} extrapolation from the infrared region to
zero frequency \cite{milton04}.
The inadequacy of neglecting the transverse momentum dependence has been
stressed by Esquivel and Svetovoy \cite{esquivel04}.

\section{Conclusions}
In this paper we have sharpened our arguments in favor of using real data
for the dielectric functions to apply the Lifshitz formula to calculate
the force between metal surfaces, in particular between a spherical lens
and a flat plate.  In principle, one can also use the surface impedance
to calculate this force, and although optical data is lacking for the latter,
such use should yield the same result.  In contrast, the procedures
advocated in Refs.~\cite{chen04,bezerra04, bezerra02,decca03,decca03a,geyer03}
contain {\it ad hoc\/} elements and assumptions.

We show both by direct computation, and through analogous models, that there
is no conflict with thermodynamical principles, in particular with the
Nernst heat theorem (the third law of thermodynamics).  Especially important
is the demonstration that the entropy necessarily vanishes at zero temperature.
Claims that experimental limits on Casimir forces preclude our temperature
dependence \cite{decca03,decca03a} are, in our opinion,
not justified, since the
accuracy of the current experiments does not match their precision, especially
due to the impossibility of determining the shortest separation distances
accurately.  Undoubtedly, it will take dedicated experiments involving 
different temperatures to reveal the true temperature dependence of the Casimir
effect.  One idea might be to measure the difference between the Casimir 
forces for the same value of $a$ at two different temperatures, for instance 
300 K and 350 K.  Such a difference is directly measurable, in principle. 
To our knowledge this idea was originally proposed by 
Chen {\it et al.}~\cite{chen03},
and  it was further discussed in Ref.~\cite{brevik04}.

\acknowledgments
I.B. thanks 
Astrid Lambrecht and Serge
Reynaud for providing their numerical results for the
permittivities of Au, Cu, and Al.  The work of K.A.M. is supported
in part by the U.S. Department of Energy.

\end{document}